\documentstyle[12pt,a4,psfig]{article}
\begin{document}
\author{
Santi Prestipino \\
Universit\`a degli Studi di Messina, Dipartimento di Fisica, and \\
Istituto Nazionale per la Fisica della Materia (INFM), \\
Contrada Papardo, 98166 Messina, Italy; \\
e-mail: {\tt Santi.Prestipino@unime.it}}
\title{The ideal gas as an urn model: \\
derivation of the entropy formula}
\maketitle
\begin{abstract}
~~The approach of an ideal gas to equilibrium is simulated through a
generalization of the Ehrenfest ball-and-box model.
In the present model, the interior of each box is discretized, {\it i.e.},
balls/particles live in cells whose occupation can be either multiple or
single.
Moreover, particles occasionally undergo random, but elastic, collisions
between each other and against the container walls.
I show, both analitically and numerically, that the number and energy of
particles in a given box eventually evolve to an equilibrium distribution
$W$ which, depending on cell occupations, is binomial or hypergeometric
in the particle number and beta-like in the energy.
Furthermore, the long-run probability density of particle velocities is
Maxwellian, whereas the Boltzmann entropy $\ln W$ exactly reproduces the
ideal-gas entropy.
Besides its own interest, this exercise is also relevant for pedagogical
purposes since it provides, although in a simple case, an explicit
probabilistic foundation for the ergodic hypothesis and for the
maximum-entropy principle of thermodynamics.
For this reason, its discussion can profitably be included in a graduate
course on statistical mechanics.
\end{abstract}

\vspace{5mm}

\vspace{2mm}

\newpage
\section{Introduction}

After the pioneering work of Boltzmann, there is now a general consensus
on the idea that a dynamically chaotic motion generically leads, in systems
of very many particles, to thermodynamic behaviour.
A general proof of this statement is however lacking, and one's intuition
usually appeals to simplified dynamical models which allow for some
analytic treatment.
Statistical toy-models that illustrate how thermodynamic equilibrium is
established in practice are especially helpful for educational scope,
since they supply students with a plain justification and a direct
understanding of the basic assumptions of thermodynamics and statistical
mechanics.
In particular, an intuitive picture of the emergence of the Second Law of
thermodynamics from mechanics is provided by the behaviour of stochastic
urn models, where balls/particles are subjected to a {\em probabilistic}
evolution which, eventually, drives the system towards a stationary
state~\cite{Ehrenfest}.
Obviously, this stochastic (Markovian) dynamics is only a caricature of
the ``real'' (Newtonian) dynamics; it is much like an effective dynamics
which emerges after averaging over many instances of the complicated
short-time motion.

In the Ehrenfest model, $N$ numbered balls are distributed into two urns;
at each time step, a number between 1 and $N$ is extracted at random, and
the ball with that label is moved from the urn where it resides to the other.
Eventually, the average number of balls in each urn becomes equal to $N/2$,
with relative fluctuations around the mean that are negligible in the
large-$N$ limit.
This stochastic process, which Kac considered as ``probably one of the
most instructive models in the whole of Physics''~\cite{Kac}, gives an
illustration of the irreversible diffusive dynamics of two
dilute gases of the same species, hosted in two communicating, but
globally isolated, vessels of equal volume.
As we learn from thermodynamics, and is confirmed by experience, the two
gases eventually attain an equilibrium state being characterized by an
equal number of particles in the two vessels.

In order to include also energetic considerations into the description,
I consider a generalization of the Ehrenfest model where the balls/particles
are endowed with both a discrete position and a continuous velocity.
To be specific, we are given two boxes, 1 and 2, and $N$ labelled
particles distributed between the boxes.
Box 1 (2) is divided into $V_1$ ($V_2$) identical cells, $V=V_1+V_2$ being
the total cell number.
The occupation number $c_{\alpha}$ of the $\alpha$-th cell
($\alpha=1,\ldots,V$) can be either Bose-like ($c_{\alpha}=0,1,2,\ldots$)
or Fermi-like ($c_{\alpha}=0,1$), with both possibilities being considered
in the following.
The velocity of the $a$-th particle ($a=1,\ldots,N$) is ${\bf v}_a$, a
three-dimensional vector with components $v_{ak},\,k=1,2,3$.

To make some progress in the analytic treatment of the model, a drastic
simplification is made, namely that the position dynamics is totally
decoupled from the velocity dynamics.
This is obtained by an independent and alternate updating of positions
and velocities, in such a way that free diffusive motion and collisions
will run in parallel, yet staying separate.
In particular, the equilibrium of one set of variables can be analysed
without making reference to the other.
The assumption of decoupling between position and velocity updating is
tantamount to the hypothesis that 1) the Markov time step, while being
much longer than any microscopic collision time, is nevertheless shorter
than the time needed for the velocity distribution to relax ({\it i.e.},
to reach equilibrium); 2) velocity relaxation occurs on a time scale
that is also well separated from (typically much longer than) the
equilibration time of the number density.

I argue, and indeed is verified {\it a posteriori}, that this model gives
a representation of the ideal-gas dynamics once defining the entropy with
the logarithm of the probability density of macrostate variables.
Hence, by this route one arrives at a novel ({\it i.e.}, not based on
the microcanonical ensemble) microscopic foundation of the ideal-gas
expression for the entropy and, at the same time, at a probabilistic
justification of the maximum-entropy principle of thermodynamics.
Furthermore, this simple model gives the opportunity to discuss at length
the issue of ergodicity of a probabilistic evolution and its relevance for
the actual deterministic dynamics of a many-particle system.
This point is usually hardly understood by graduate students in statistical
physics, who find it rather obscure.
The present model could come in useful for providing an easy access to such
basic theoretical questions.

The outline of this paper is the following: in Sections 2 and 4, I describe
the stochastic dynamics of particle positions and velocities, respectively.
Section 3 is an {\it intermezzo}, mainly of illustrative value, where I show
an example of exact derivation of a macro-variable (coarse-grained) evolution
from the microscopic dynamics.
Further comments and conclusions are given in Section 5.

\section{Update of positions}

Let us first suppose that each cell in the boxes can host whatever number
of particles.
A positions update consists of 1) choosing at random one particle, $a_r$,
and one cell, $\alpha_r$; and 2) moving particle $a_r$ into cell $\alpha_r$.
In terms of the macro-variable $n$, which counts how many particles are
currently found in box 1, this defines a stationary stochastic process of
the Markov type, being characterized by the following conditional
(or transition) probabilities:
\begin{eqnarray}
T(n+1\leftarrow n) &\equiv& P(n+1;t+1|n;t)=\frac{(N-n)V_1}{NV}\,;
\nonumber \\
T(n-1\leftarrow n) &\equiv& P(n-1;t+1|n;t)=\frac{nV_2}{NV}\,.
\label{eq01}
\end{eqnarray}
In Eq.\,(\ref{eq01}), $t=0,1,2,\ldots$ is a discrete time.
The ensuing master equation for $n$ reads:
\begin{eqnarray}
P(n;t+1) &=& \frac{(N-n+1)V_1}{NV}P(n-1;t)+\frac{(n+1)V_2}{NV}P(n+1;t)
\nonumber \\
&+& \left( 1-\frac{(N-n)V_1}{NV}-\frac{nV_2}{NV}\right) P(n;t)\,.
\label{eq02}
\end{eqnarray}

It is immediate to realize, by direct inspection, that Eq.\,(\ref{eq02})
admits the {\em binomial} distribution
\begin{equation}
W(n)={N\choose n}\left( \frac{V_1}{V}\right) ^n
\left( \frac{V_2}{V}\right) ^{N-n}\,
\label{eq03}
\end{equation}
as unique stationary distribution.
Since the Markov chain is ergodic ({\it i.e.}, there is a path connecting
every (macro)state $n$ to every other $n'$), any initial distribution
$P(n;0)$ will converge, in the long run, to $W(n)$.
Observe that $W(n)$ gives the chance that, upon randomly distributing
$N$ labelled particles into two boxes, with different {\it a priori}
probabilities for the boxes, the number of particles in box 1 be $n$.
Moreover, the multiplicity of macrostate $n$, {\it i.e.}, the number of
complexions (microstates) of $N$ distinguishable particles in the boxes,
such that box 1 contains $n$ particles, is $V^NW(n)$.
Hereafter, I list a number of properties which hold for the dynamics
ruled by Eq.\,(\ref{eq02}).

First, I calculate the average $n$ and $n^2$ at any time by explicitly
evaluating the two sums
\begin{eqnarray}
\left< n\right> (t+1) &\equiv& \sum_{n=0}^NnP(n;t+1)=\frac{V_1}{V}+
\left( 1-\frac{1}{N}\right) \left< n\right> (t)\,;
\nonumber \\
\left< n^2\right> (t+1) &\equiv& \sum_{n=0}^Nn^2P(n;t+1)
\nonumber \\
&=& \frac{V_1}{V}+\left( 1-\frac{2}{N}\right) \left< n^2\right> (t)+
\left[\left( 1-\frac{1}{N}\right) \frac{2V_1}{V}+\frac{1}{N}\right]
\left< n\right> (t)\,.
\label{eq04}
\end{eqnarray}
The first of the difference equations (\ref{eq04}) admits a solution
$\left< n\right> (t)=a+bx^t$, for suitable $a,b$, and $x$.
One easily finds:
\begin{equation}
\left< n\right> (t)=\frac{NV_1}{V}+\left( \left< n\right> (0)-
\frac{NV_1}{V}\right) \left( 1-\frac{1}{N}\right) ^t\,,
\label{eq05}
\end{equation}
{\it i.e.}, an irreversible exponential approach to $NV_1/V$, a value
corresponding to having the same density of particles in every box.
Similarly, the second of Eqs.\,(\ref{eq04}) has a solution of the form
\begin{equation}
\left< n^2\right> (t)=a'+b'\left( 1-\frac{1}{N}\right) ^t+
c'\left( 1-\frac{2}{N}\right) ^t\,,
\label{eq06}
\end{equation}
with $a'=NV_1/V+N(N-1)(V_1/V)^2$.
Whence the variance of $n$ scales, in the infinite-time limit, as $N$:
\begin{equation}
\frac{\sqrt{\left< n^2\right> -\left< n\right> ^2}}{\left< n\right>}
\rightarrow \frac{1}{\sqrt{N}}\sqrt{\frac{V_2}{V_1}}\,.
\label{eq07}
\end{equation}

Assuming $n,N-n,V_1,V_2={\cal O}(N)\gg 1$ and using the Stirling
approximation $\ln N!=N(\ln N-1)+{\cal O}(\ln N)$, the equilibrium
entropy $S(n)$, defined as the logarithm of the multiplicity $V^NW(n)$,
turns out to be additive over the boxes and extensive with $n$:
\begin{equation}
S(n)\sim -n\ln\frac{n}{V_1}-(N-n)\ln\frac{N-n}{V_2}+N\ln N\,,
\label{eq08}
\end{equation}
being maximum for $n=NV_1/V$ (subextensive terms in (\ref{eq08}) are ignored).
Note that the superextensive $N\ln N$ term turns into an extensive constant
if $V^NW(n)$ is multiplied by the Gibbs factor of $1/N!$, whose origin is
quantum-mechanical (it arises as a required correction to the partition
function of a system of identical particles in the classical limit,
{\it i.e.}, under the hypothesis of low particle density at {\it all}
temperatures).
In Eq\,(\ref{eq08}), we recognize the volume contribution to the ideal-gas
entropy.
Therefore, the equilibrium (and asymptotic) value of $n$ is the outcome of
the entropy maximization, as indeed prescribed by thermodynamics.

It easily follows from Eq.\,(\ref{eq03}) that the profile of $W(n)$ around
the maximum is Gaussian,
\begin{equation}
\ln W(n)={\rm const.}-\frac{V^2\Delta n^2}{2NV_1V_2}+{\cal O}\left(
\frac{\Delta n^3}{N^2}\right) \,,
\label{eq09}
\end{equation}
$\Delta n=n-NV_1/V$ being the deviation from the abscissa of the maximum.
The last term in Eq.\,(\ref{eq09}) is negligible for standard deviations
$\Delta n={\cal O}(\sqrt N)$.

Nothing changes in the asymptotics if $\alpha_r$ is forbidden to be the
same original cell of $a_r$.
In this case, 
\begin{equation}
T(n+1\leftarrow n)=\frac{(N-n)V_1}{N(V-1)}\,;\,\,\,\,\,\,
T(n-1\leftarrow n)=\frac{nV_2}{N(V-1)}\,,
\label{eq10}
\end{equation}
but the form of $W(n)$ is unchanged.
However,
\begin{equation}
\left< n\right> (t)=\frac{NV_1}{V}+\left( \left< n\right> (0)-
\frac{NV_1}{V}\right) \left( 1-\frac{V}{N(V-1)}\right) ^t
\label{eq11}
\end{equation}
is slightly different from (\ref{eq05}), though sharing the same limit
$NV_1/V$ for $t\rightarrow\infty$.

While all of the above sounds quite ``standard'', novel results are those
I obtain for the case of single-occupation cells.
Now, at each step of the process, the selected particle $a_r$ is moved
into a cell $\alpha_r$ that is chosen at random among the vacant sites.
The transition probabilities now read (with $V_1,V_2\geq N$):
\begin{equation}
T(n+1\leftarrow n)=\frac{(N-n)(V_1-n)}{N(V-N)}\,;\,\,\,\,\,\,
T(n-1\leftarrow n)=\frac{n(V_2-N+n)}{N(V-N)}\,,
\label{eq12}
\end{equation}
yielding a {\em hypergeometric} stationary distribution for variable $n$:
\begin{equation}
W(n)={V\choose N}^{-1}{V_1\choose n}{V_2\choose N-n}\,.
\label{eq13}
\end{equation}
The Vandermonde identity,
\begin{equation}
\sum_{n=0}^N{V_1\choose n}{V_2\choose N-n}={V_1+V_2\choose N}\,,
\label{eq14}
\end{equation}
ensures that Eq.\,(\ref{eq13}) is normalized correctly.
$W(n)$ gives the chance that, upon randomly choosing $N$ cells ({\it i.e.},
the occupied ones) among a total of $V$ distinguishable sites, the number
of particles in box 1 be $n$.
Stated it differently, ${N\choose n}$ is the number of ways $n$ numbered
particles can be sorted out from a set of $N$, while
\begin{equation}
\frac{V_1}{V}\,\frac{V_1-1}{V-1}\,\cdots\frac{V_1-(n-1)}{V-(n-1)}\,
\frac{V_2}{V-n}\,\frac{V_2-1}{V-(n+1)}\,\cdots\frac{V_2-(N-n-1)}{V-(N-1)}
\label{eq15}
\end{equation}
is the number of ways {\em these} $n$ particles can be allocated in box 1
(the other $N-n$ being attributed to box 2 instead).
The product of ${N\choose n}$ by (\ref{eq15}) gives again $W(n)$.
Finally, the multiplicity of state $n$, {\it i.e.}, the number of ways $N$
indistinguishable particles can be arranged into $V$ distinguishable cells,
in such a way that $n$ particles reside in box 1, is equal to
${V_1\choose n}{V_2\choose N-n}={V\choose N}W(n)$.

The average $n$ and $n^2$ at time $t$ are given by:
\begin{eqnarray}
\left< n\right> (t) &=& \frac{NV_1}{V}+
\left( \left< n\right> (0)-\frac{NV_1}{V}\right)
\left( 1-\frac{V}{N(V-N)}\right) ^t\,;
\nonumber \\
\left< n^2\right> (t) &=& a'+b'\left( 1-\frac{V}{N(V-N)}\right) ^t+
c'\left( 1-\frac{2(V-1)}{N(V-N)}\right) ^t\,,
\label{eq16}
\end{eqnarray}
with $a'=\left[ NV_1V_2+N^2V_1(V_1-1)\right] /\left[ V(V-1)\right] $.
In the infinite-time limit, the relative deviation from the average
\begin{equation}
\frac{\sqrt{\left< n^2\right> -\left< n\right> ^2}}{\left< n\right>}
\rightarrow\frac{1}{\sqrt{N}}\sqrt{\frac{V_2(V-N)}
{V_1(V-1)}}\,.
\label{eq17}
\end{equation}

Upon assuming $n,N-n,V_1-n,V_2-N+n={\cal O}(N)\gg 1$, the equilibrium
entropy becomes:
\begin{eqnarray}
S(n)\equiv\ln\left[ {V_1\choose n}{V_2\choose N-n}\right] &\sim&
-n\ln\frac{n}{V_1}-(V_1-n)\ln\left( 1-\frac{n}{V_1}\right)
\nonumber \\
&-& (N-n)\ln\frac{N-n}{V_2}-(V_2-N+n)\ln\left( 1-\frac{N-n}{V_2}\right) \,,
\label{eq18}
\end{eqnarray}
being maximum for $n=NV_1/V$.
Equation (\ref{eq18}) is nothing but the thermodynamic entropy of two
ideal {\em lattice} gases that can mutually exchange energy and particles.

Finally, it immediately follows from Eq.\,(\ref{eq13}) that the profile of
$W(n)$ around the maximum is Gaussian,
\begin{equation}
\ln W(n)={\rm const.}-\frac{V^3\Delta n^2}{2N(V-N)V_1V_2}+
{\cal O}\left( \frac{\Delta n^3}{N^2}\right) +
{\cal O}\left( \frac{\Delta n^3}{(V-N)^2}\right) \,,
\label{eq19} 
\end{equation}
$\Delta n=n-NV_1/V$ being the deviation from the abscissa of the maximum.
Again, the last two terms in Eq.\,(\ref{eq19}) are negligible for standard
deviations $\Delta n={\cal O}(\sqrt N)$.

\section{Derivation of a coarse-grained dynamics \\
from the microstate dynamics}

For a specific instance of stochastic dynamics of cell occupation numbers,
I provide in this paragraph the detailed derivation of the coarse-grained
evolution of a macro-variable ({\it i.e.}, the number $n$ of occupied
cells in box 1).

In the example considered here, the occupation numbers $c_{\alpha}=0,1$
are made to evolve according to the following rules: at each time step,
1) two cells are chosen at random (either in the same box or in different
boxes), and 2) their occupation numbers are mutually -- and unconditionally
-- exchanged (observe that the overall number $N$ of occupied cells is
conserved by the dynamics).
The ensuing Monte Carlo/Markovian evolution is thus specified by the
transition probabilities
\begin{equation}
\tau(\{c'\}\leftarrow\{c\})=\frac{2}{V(V-1)}
\sum_{\alpha<\beta}\left( \delta_{c'_{\alpha},\,c_{\beta}}
\delta_{c'_{\beta},\,c_{\alpha}}\prod_{\gamma\neq\alpha,\,\beta}
\delta_{c'_{\gamma},\,c_{\gamma}}\right) \,,
\label{eq20}
\end{equation}
where the constant prefactor in (\ref{eq20}) ensures the correct
normalization, namely that
\begin{equation}
\sum_{\{c'\}}\tau(\{c'\}\leftarrow\{c\})=1\,.
\label{eq21}
\end{equation}
Taken $\pi(\{c\};t)$ to be the probability for the occurrence of the
microstate $\{c\}$ at time $t$, the master equation of micro-evolution
formally reads:
\begin{equation}
\pi(\{c'\};t+1)=\sum_{\{c\}}\tau(\{c'\}\leftarrow\{c\})\,\pi(\{c\};t)\,.
\label{eq22}
\end{equation}
Upon plugging Eq.\,(\ref{eq20}) into Eq.\,(\ref{eq22}), the latter
equation becomes:
\begin{equation}
\pi(\ldots,c_{\alpha},\ldots,c_{\beta},\ldots;t+1)=
\frac{2}{V(V-1)}\sum_{\alpha<\beta}
\pi(\ldots,c_{\beta},\ldots,c_{\alpha},\ldots;t)\,,
\label{eq23}
\end{equation}
which admits the constant ${V\choose N}^{-1}$ as a stationary solution.

Let the $V$ cells be numbered in such a way that the first $V_1$ cells
in the list do belong to box 1, while those from $V_1+1$ to $V$ belong
to box 2.
The probability of observing the macrostate $n$ at time $t$ is then
\begin{equation}
P(n;t)=
\sum_{\{c\}}\delta_{\sum_{\gamma=1}^{V_1}c_{\gamma},\,n}\,\pi(\{c\};t)\,.
\label{eq24}
\end{equation}
I aim at finding an equation of evolution for this $P$, namely a master
equation that is valid at a less fundamental, coarse-grained level of
description.

First, I note that every sum over all distinct pairs of cells (like that
appearing on the r.h.s. of Eq.\,(\ref{eq23})) can be decomposed into
three sums, $\sum_{\alpha<\beta\le V_1}+\sum_{\alpha\le V_1<\beta}+
\sum_{V_1<\alpha<\beta}$, the three partial sums being denoted as $A,B$,
and $C$, respectively.
Also observe that, for $\alpha<\beta\le V_1$ or $V_1<\alpha<\beta$, the
value of $n$ is left unchanged by the exchange of $c_{\alpha}$ and
$c_{\beta}$, and the same happens for $\alpha\le V_1<\beta$ provided
$c_{\alpha}=c_{\beta}$.
Conversely, for $\alpha\le V_1<\beta$, $n$ increases (or decreases) by 1
when $c_{\alpha}=0$ and $c_{\beta}=1$ (or the other way around).

Let $S_n$ be the set of all the ${V_1\choose n}{V_2\choose N-n}$
microstates $\{c\}$ such that $\sum_{\gamma=1}^{V_1}c_{\gamma}=n$.
Upon summing the l.h.s. of Eq.\,(\ref{eq23}) over $S_n$, the net
result is, by definition, $P(n;t+1)$.
Similarly, summing $A$ and $C$ over the same microstates gives
\begin{equation}
\frac{V_1(V_1-1)}{2}P(n;t)\,\,\,\,\,\,{\rm and}\,\,\,\,\,\,
\frac{V_2(V_2-1)}{2}P(n;t)\,,
\label{eq25}
\end{equation}
respectively.
As to the pairs $\alpha<\beta$ contributing to $B$, {\it i.e.}, satisfying
$\alpha\le V_1<\beta$, the microstates of $S_n$ are classified in four
categories, according to the values of $c_{\alpha}$ and $c_{\beta}$.
Calling ${\cal N}_{\alpha,\,\beta}$ the total number of pairs of each
type, one has:
\begin{eqnarray}
c_{\alpha}=c_{\beta}=1 &:& \,\,\,\,{\cal N}_{\alpha,\,\beta}^{(1)}=n(N-n)\,;
\nonumber \\
c_{\alpha}=c_{\beta}=0 &:& \,\,\,\,{\cal N}_{\alpha,\,\beta}^{(2)}=(V_1-n)(V_2-N+n)\,;
\nonumber \\
c_{\alpha}=1\,\,{\rm and}\,\,c_{\beta}=0 &:& \,\,\,\,{\cal N}_{\alpha,\,\beta}^{(3)}=n(V_2-N+n)\,;
\nonumber \\
c_{\alpha}=0\,\,{\rm and}\,\,c_{\beta}=1 &:& \,\,\,\,{\cal N}_{\alpha,\,\beta}^{(4)}=(V_1-n)(N-n)\,.
\label{eq26}
\end{eqnarray}
In particular, the coefficient of $P(n;t)$ in the master equation for $n$
should be:
\begin{eqnarray}
&& \frac{2}{V(V-1)}\left( \frac{V_1(V_1-1)}{2}+\frac{V_2(V_2-1)}{2}+
n(N-n)+(V_1-n)(V_2-N+n)\right)
\nonumber \\
&=& 1-\frac{2(V_1-n)(N-n)+2n(V_2-N+n)}{V(V-1)}\,.
\label{eq27}
\end{eqnarray}

Now observe that, for each microstate of $S_n$, there are exactly
${\cal N}_{\alpha,\,\beta}^{(3)}$ terms in the sum
$\sum_{\alpha\le V_1<\beta}\pi(\ldots,c_{\beta},\ldots,c_{\alpha},\ldots;t)$
that refer to microstates of $S_{n-1}$.
Hence, the coefficient of $P(n-1;t)$ in the master equation ({\it i.e.},
the number of times any specific microstate of $S_{n-1}$ is repeated in
the sum $\sum_{\{c\}\in S_n}\sum_{\alpha\le V_1<\beta}
\pi(\ldots,c_{\beta},\ldots,c_{\alpha},\ldots;t)$) is given by the product
of ${\cal N}_{\alpha,\,\beta}^{(3)}$ times the number of $S_n$ microstates,
divided by the total number of $S_{n-1}$ microstates:
\begin{equation}
\frac{n(V_2-N+n){V_1\choose n}{V_2\choose N-n}}
{{V_1\choose n-1}{V_2\choose N-n+1}}=(V_1-n+1)(N-n+1)\,,
\label{eq28}
\end{equation}
which is also the number of $S_n$ microstates that can be originated from
any specific microstate of $S_{n-1}$ by moving a particle from box 2 to 1.

A similar calculation for the coefficient of $P(n+1;t)$ yields:
\begin{equation}
\frac{(V_1-n)(N-n){V_1\choose n}{V_2\choose N-n}}
{{V_1\choose n+1}{V_2\choose N-n-1}}=(n+1)(V_2-N+n+1)\,.
\label{eq29}
\end{equation}
In the end, the complete master equation for $n$ reads:
\begin{eqnarray}
P(n;t+1) &=& \frac{2(V_1-n+1)(N-n+1)}{V(V-1)}P(n-1;t)
\nonumber \\
&+& \frac{2(n+1)(V_2-N+n+1)}{V(V-1)}P(n+1;t)
\nonumber \\
&+& \left( 1-\frac{2(V_1-n)(N-n)+2n(V_2-N+n)}{V(V-1)}\right) P(n;t)\,.
\label{eq30}
\end{eqnarray}
One can easily extract from the above equation the expression of the
transition probabilities, with the result:
\begin{equation}
T(n+1\leftarrow n)=\frac{2(V_1-n)(N-n)}{V(V-1)}\,;\,\,\,\,\,\,
T(n-1\leftarrow n)=\frac{2n(V_2-N+n)}{V(V-1)}\,.
\label{eq31}
\end{equation}
The latter probabilities, although not identical to the (\ref{eq12}),
nonetheless lead to the same stationary distribution (\ref{eq13}), as
can be checked directly.
This is not strange, since the transition probabilities (\ref{eq31})
are obtained by multiplying the (\ref{eq12}) for the constant factor
$2N(V-N)/[V(V-1)]$.

\section{Update of velocities}

The collision dynamics of a set of equal-mass particles can be
schematized, at the roughest level of description, as a succession
of {\em random} binary events which are nevertheless required to obey
energy and momentum conservation~\cite{Sauer}.
On the macroscopic side, such collision rules go along with the
conservation of total {\em kinetic} energy and total momentum, thus
being appropriate only to a very dilute (gaseous) system of particles.
If, moreover, we want to drop the momentum constraint, provision should
be made also for elastic collisions of particles against the (cubic)
container walls, causing the reversal of one component only of the
velocity of the hitting particle (say, the $x$ component if the collision
occurs against the wall that is orthogonal to the $x$ axis).

As far as the mutual collisions are concerned, the conservation laws by
themselves require that the velocities of the colliding particles, say
$a$ and $b$, be updated as:
\begin{equation}
{\bf v}_a\rightarrow {\bf v}'_a=
{\bf v}_a+(\Delta v)\hat{\bf r}\,;\,\,\,\,\,\,
{\bf v}_b\rightarrow {\bf v}'_b=
{\bf v}_b-(\Delta v)\hat{\bf r}\,,
\label{eq32}
\end{equation}
where $\Delta v=({\bf v}_b-{\bf v}_a)\cdot\hat{\bf r}$, and all that we
know about the unit-length vector $\hat{\bf r}$ is that it forms an acute
angle with ${\bf v}_b-{\bf v}_a$ ({\it i.e.}, $\Delta v>0$).
In particular, note that a general property of an elastic collision is:
\begin{equation}
\left| {\bf v}'_a-{\bf v}'_b\right| =\left| {\bf v}_a-{\bf v}_b\right| \,.
\label{eq33}
\end{equation}
The full specification of $\hat{\bf r}$ would require more knowledge
about the collision ({\it i.e.}, the peculiar geometry of the impact
and the exact law of interaction between the particles).
Instead, the collision rules considered here are such that the outcome
of a mutual collision is {\em as maximally random as possible}: that is,
at each step of the game, $\hat{\bf r}$ is picked up at random from the
emisphere of unit vectors forming an acute angle with
${\bf v}_b-{\bf v}_a$~\cite{note}.
Note that in one dimension only is the vector $\hat{\bf r}$ nonetheless
univocally determined: this is consistent with the known fact that, for
particles moving on a straight line and colliding elastically, the
conservation laws suffice to determine the post-collision velocities
from the initial ones.
I point out that the duration of velocity relaxation in a real system is
rather sensitive to the peculiarities of the interaction between particles.
However, this may not be the case for the asymptotic shape of the velocity
distribution, which is only aware of the conservation laws that rule the
outcome of an individual collision.

If collisions against walls and between particles occur at a rate of
$1-p$ and $p$, respectively (where $p$ is any number between 0 and 1),
the master equation for the velocities reads:
\begin{equation}
\pi(\{{\bf v}'\};t+1)=\int{\rm d}^{3N}v\,
\tau(\{{\bf v}'\}\leftarrow\{{\bf v}\})\pi(\{{\bf v}\};t)\,,
\label{eq34}
\end{equation}
where $\tau=(1-p)\tau_1+p\tau_2$ and
\begin{eqnarray}
\tau_1(\{{\bf v}'\}\leftarrow\{{\bf v}\}) &=& \frac{1}{3N}
\sum_{a=1}^N\sum_{k=1}^3\left[ \delta(v'_{ak}+v_{ak})
\prod_{(b,\,l)\neq(a,\,k)}\delta(v'_{bl}-v_{bl})\right] \,;
\nonumber \\
\tau_2(\{{\bf v}'\}\leftarrow\{{\bf v}\}) &=&
\frac{2}{N(N-1)}\sum_{a<b}\left[
\frac{1}{2\pi\left| {\bf v}_a-{\bf v}_b\right| }
\delta^3({\bf v}'_a+{\bf v}'_b-{\bf v}_a-{\bf v}_b)
\delta(v_a^{\prime 2}+v_b^{\prime 2}-v_a^2-v_b^2)\right.
\nonumber \\
&\times&\left.
\prod_{c\neq a,\,b}\delta^3({\bf v}'_c-{\bf v}_c)\right] \,.
\label{eq35}
\end{eqnarray}
Note that the two kernels $\tau_1$ and $\tau_2$ are separately normalized.
In particular, the factor $1/(2\pi\left| {\bf v}_a-{\bf v}_b\right| )$ is
the outcome of a six-dimensional integration of delta functions, which is
performed by substituting ${\bf v}'_a$ and ${\bf v}'_b$ with the auxiliary
variables
${\bf s}=({\bf v}'_a+{\bf v}'_b)/2$ and ${\bf t}=({\bf v}'_a-{\bf v}'_b)/2$
(the jacobian for this transformation is 8):
\begin{eqnarray}
&& \int{\rm d}^3v'_a\,{\rm d}^3v'_b\,%
\delta^3({\bf v}'_a+{\bf v}'_b-{\bf v}_a-{\bf v}_b)\,%
\delta(v_a^{\prime 2}+v_b^{\prime 2}-v_a^2-v_b^2)
\nonumber \\
&=& 8\int{\rm d}^3s\,{\rm d}^3t\,%
\delta^3[2{\bf s}-({\bf v}_a+{\bf v}_b)]\,%
\delta[2(s^2+t^2)-(v_a^2+v_b^2)]
\nonumber \\
&=& 8\pi\int_0^{+\infty}{\rm d}t\,t^2
\delta\left[ t^2-\left( \frac{{\bf v}_a-{\bf v}_b}{2}\right) ^2\right]
\nonumber \\
&=& 2\pi\left| {\bf v}_a-{\bf v}_b\right| \,.
\label{eq36}
\end{eqnarray}

Using Eqs.\,(\ref{eq33}) and (\ref{eq36}), it is easy to prove that a
stationary solution to Eq.\,(\ref{eq34}) is, for any regular and properly
normalized function $F$:
\begin{equation}
w(\{{\bf v}\})=F(v_1^2+\cdots+v_N^2)\,.
\label{eq37}
\end{equation}
An outstanding exception is $p=1$, where the more general stationary
solution to Eq.\,(\ref{eq34}) is instead
$F(v_1^2+\cdots+v_N^2)\,G({\bf v}_1+\cdots+{\bf v}_N)$, for arbitrary
$F$ and $G$ functions.

Next, I consider the one- and two-body velocity distributions at time $t$.
These are marginal distributions that are built over $\pi(\{{\bf v}\};t)$:
\begin{eqnarray}
f_1({\bf v}_1;t) &=&
\int{\rm d}^3v_2{\rm d}^3v_3\ldots{\rm d}^3v_N\,\pi(\{{\bf v}\};t)\,;
\nonumber \\
f_2({\bf v}_1,{\bf v}_2;t) &=&
\int{\rm d}^3v_3\ldots{\rm d}^3v_N\,\pi(\{{\bf v}\};t)\,.
\label{eq38}
\end{eqnarray}
Seeking for an exact equation of evolution for $f_1$, I calculate
$f_1({\bf v}_1;t+1)$ by inserting Eq.\,(\ref{eq34}) into the first of
Eqs.\,(\ref{eq38}).
While the term arising from $\tau_1$ can be easily worked out, less
straightforward is the derivation of the other one, involving $\tau_2$:
\begin{eqnarray}
&& \int{\rm d}^3v_2\ldots{\rm d}^3v_N\,
\int{\rm d}^{3N}v'\,\tau_2(\{{\bf v}\}\leftarrow\{{\bf v}'\})
\pi(\{{\bf v}'\};t)
\nonumber \\
&=& \left( 1-\frac{2}{N}\right) f_1({\bf v}_1;t)+\frac{2}{N(N-1)}\sum_{b>1}
\int{\rm d}^{3N}v'\,\frac{1}{2\pi|{\bf v}'_1-{\bf v}'_b|}\pi(\{{\bf v}'\};t)
\nonumber \\
&\times&
\int{\rm d}^3v_b\,\delta^3({\bf v}_1+{\bf v}_b-{\bf v}'_1-{\bf v}'_b)\,%
\delta(v_1^2+v_b^2-v_1^{\prime 2}-v_b^{\prime 2})\,,
\label{eq39}
\end{eqnarray}
where the latter sum is actually made of $N-1$ identical contributions.
Then, considerations similar to those leading to Eq.\,(\ref{eq36}) allow
one to further simplify the r.h.s. of Eq.\,(\ref{eq39}) and to arrive at
the final equation for $f_1$, whose status is akin to that of the famous
Boltzmann equation in the kinetic theory of gases:
\begin{eqnarray}
f_1({\bf v}_1;t+1) &=&
(1-p)\left\{ \left( 1-\frac{1}{N}\right) f_1({\bf v}_1;t)\right.
\nonumber \\
&+& \left. \frac{1}{3N}\left[ f_1(-v_{1x},v_{1y},v_{1z};t)+
f_1(v_{1x},-v_{1y},v_{1z};t)+f_1(v_{1x},v_{1y},-v_{1z};t)\right] \right\}
\nonumber \\
&+& p\left\{ \left( 1-\frac{2}{N}\right) f_1({\bf v}_1;t)
+\frac{2}{N}\times\frac{1}{2\pi}\int{\rm d}^3v_2\int{\rm d}^3\Delta\,
\frac{1}{\Delta}\,\delta\left[
\Delta^2-\left( \frac{{\bf v}_1-{\bf v}_2}{2}\right) ^2\right] \right.
\nonumber \\
&\times& \left. f_2\left( \frac{{\bf v}_1+{\bf v}_2}{2}+
{\bf\Delta},\frac{{\bf v}_1+{\bf v}_2}{2}-{\bf\Delta};t\right) \right\} \,.
\label{eq40}
\end{eqnarray}
While it seems problematic to derive a sort of H-theorem from
Eq.\,(\ref{eq40}), a more easy program to fulfil is to find
time-independent solutions for this equation.
It is immediate to check that the
{\it ansatz} $f_2^{\rm (eq)}({\bf v}_1,{\bf v}_2)=\Phi(v_1^2+v_2^2)$
gives a stationary solution to Eq.\,(\ref{eq40}) for any appropriate
function $\Phi$ (for $p=1$, the more general time-independent solution
is instead $\Phi(v_1^2+v_2^2)\Psi({\bf v}_1+{\bf v}_2)$).
However, in case of an isolated system with total energy $U$~\cite{note2},
the only admissible solution (\ref{eq37}) is the microcanonical density
\begin{equation}
w(\{{\bf v}\})=\frac{\Gamma(3N/2)}{\pi^{3N/2}}
U^{-\left( \frac{3N}{2}-1\right) }\delta(v_1^2+\cdots +v_N^2-U)\,.
\label{eq41}
\end{equation}
In this case, the $\Phi$ function can be explicitly worked out by
transforming to hyperspherical coordinates:
\begin{eqnarray}
f_2^{\rm (eq)}({\bf v}_1,{\bf v}_2) &=&
\frac{\int{\rm d}^3v_3\ldots{\rm d}^3v_N\,
\delta(U-v_1^2-v_2^2-\sum_{a=3}^Nv_a^2)}
{\int{\rm d}^3v_1\ldots{\rm d}^3v_N\,\delta(U-\sum_{a=1}^Nv_a^2)}
\nonumber \\
&=& \frac{S_{3(N-2)}(1)\int_0^{+\infty}{\rm d}r\,r^{3(N-2)-1}
\delta[r^2-(U-v_1^2-v_2^2)]}
{S_{3N}(1)\int_0^{+\infty}{\rm d}r\,r^{3N-1}\delta(r^2-U)}
\nonumber \\
&=& \frac{\Gamma(3N/2)}{\Gamma(3(N-2)/2)}(\pi U)^{-3}
\left( 1-\frac{v_1^2+v_2^2}{U}\right) ^{\frac{3(N-2)}{2}-1}\,,
\label{eq42}
\end{eqnarray}
$S_n(R)=2\pi^{n/2}R^{n-1}/\Gamma(n/2)$ being the area of the
$n$-dimensional hyperspherical surface of radius $R$.
A similar calculation leads to:
\begin{equation}
f_1^{\rm (eq)}({\bf v}_1)=
\frac{\Gamma(3N/2)}{\Gamma(3(N-1)/2)}(\pi U)^{-\frac{3}{2}}
\left( 1-\frac{v_1^2}{U}\right) ^{\frac{3(N-1)}{2}-1}\,,
\label{eq43}
\end{equation}
which is the finite-$N$ Maxwell-Boltzmann (MB) distribution~\cite{Mello}.
In the $N,U\rightarrow\infty$ limit (with $U/N={\cal O}(1)$), one recovers
from Eq.\,(\ref{eq43}) the more familiar Gaussian form:
\begin{equation}
f_1^{\rm (eq)}({\bf v})=\left( \frac{\kappa}{\pi}\right) ^{\frac{3}{2}}
e^{-\kappa v^2}\,,
\label{eq44}
\end{equation}
with $\kappa=3N/(2U)$, corresponding to an average $v_a^2$ of $U/N$
for all $a$.
Note that full independence of ${\bf v}_1$ and ${\bf v}_2$, namely
$f_2^{\rm (eq)}({\bf v}_1,{\bf v}_2)=
f_1^{\rm (eq)}({\bf v}_1)f_1^{\rm (eq)}({\bf v}_2)$, requires the
thermodynamic limit $N\rightarrow\infty$ and $U={\cal O}(N)$.

As a further comment, I emphasize that a distribution like (\ref{eq37})
is a meaningful solution to Eq.\,(\ref{eq34}) also for $F$ not being a
delta function.
In fact, the $t=0$ velocity distribution need not necessarily correspond
to a single microstate or to a mixture of microstates all having the same
energy $U$.
It is an equally valid possibility that the initial state encompasses a
whole distribution of microstate energies.
In this case, and taken for granted that the evolution is ergodic, the
collisions will eventually suppress any difference in weight between
the microstates having the same energy, but preserving the overall
frequency of occurrence of every energy value in the mixture.

A normalized $w$ distribution of the form (\ref{eq37}) requires that
$F$ satisfies
\begin{equation}
\int_0^{+\infty}{\rm d}U\,F(U)U^{\frac{3N}{2}-1}=
\frac{\Gamma(3N/2)}{\pi^{3N/2}}\,.
\label{eq45}
\end{equation}
Upon observing that
\begin{equation}
F(v_1^2+\cdots+v_N^2)=
\int_0^{+\infty}{\rm d}U\,F(U)\delta(v_1^2+\cdots+v_N^2-U)\,,
\label{eq46}
\end{equation}
the two- and one-body velocity distributions will read:
\begin{eqnarray}
f_2^{\rm (eq)}({\bf v}_1,{\bf v}_2) &=&
\frac{\pi^\frac{3(N-2)}{2}}{\Gamma(3(N-2)/2)}
\int_{v_1^2+v_2^2}^{+\infty}{\rm d}U\,
F(U)(U-v_1^2-v_2^2)^{\frac{3(N-2)}{2}-1}\,;
\nonumber \\
f_1^{\rm (eq)}({\bf v}_1) &=&
\frac{\pi^\frac{3(N-1)}{2}}{\Gamma(3(N-1)/2)}
\int_{v_1^2}^{+\infty}{\rm d}U\,
F(U)(U-v_1^2)^{\frac{3(N-1)}{2}-1}\,.
\label{eq47}
\end{eqnarray}
The above distributions have not generally a Gaussian profile, even
in the thermodynamic limit (an exception is $F(U)=\pi^{-3N/2}\exp(-U)$,
which leads to $f_1^{\rm (eq)}({\bf v}_1)=\pi^{-3/2}\exp(-v_1^2)$ and
$f_1^{\rm (eq)}({\bf v}_1,{\bf v}_2)=\pi^{-3}\exp(-v_1^2-v_2^2)$).

I have carried out a computer simulation of the evolution encoded in
Eq.\,(\ref{eq34}) in order to check whether the stationary distribution
(\ref{eq43}) is also an asymptotic, $t\rightarrow\infty$ solution to
Eq.\,(\ref{eq34}), as one may surmise (at least for $0<p<1$) from the
likely ergodic character of its kernel $\tau$.
First, I set $N=3$ and $U=0.06$, with $p=0.5$ (note that the choice of
$U$ is rather immaterial, it just sets the range of fluctuations of a
single velocity component to approximately a value of $2\sqrt{U/(3N)}$).
Starting from a system of velocities in any particular microstate of
energy $U$, I collect in a hystogram the values, at regular time intervals,
of the three components of, say, the velocity of particle 1.
A look at Fig.\,1 indeed shows that this hystogram has, in the long run,
the finite-$N$ MB form.
This is indirect evidence that the simulation trajectory samples uniformly,
at least effectively if not literally, the $3N$-dimensional hypersurface
of energy $U$.
I note that ergodicity does not hold for $p=0$ ({\it i.e.}, when collisions
against the walls are the only present), whereas the stochastic evolution
for $p=1$ ({\it i.e.}, only mutual collisions present) retains memory of
the initial value of the total momentum.

Afterwards, I take $N=1000$ and $U=20$ ({\it i.e.}, same $U/N$ as in the
case before), and follow the evolution of
the same hystogram as above, now starting from velocity values that
are randomly extracted from {\it e.g.} a (bounded) {\em uniform}
one-particle distribution of zero average and variance equal to $U/(3N)$
(I have checked that nothing changes in the results if the shape of the
initial one-velocity distribution were different, {\it e.g.} truncated
quadratic).
After discarding the initial part of the simulation trajectory, the
long-run distribution of values for velocity no.\,1 now compares well
with a Gaussian (see Fig.\,2), that is with the large-$N$ form of the
MB distribution.
In fact, also the instantaneous velocities of all particles are
asymptotically distributed, for large $N$, according to the same
Gaussian (see Fig.\,3).
This indicates that: 1) the vast majority of points in the energy
hypersurface is made of ``typical'' states, {\it i.e.}, microstates
that look more or less similar as far as low-order, marginal
distributions like $f_1$ are concerned; 2) the microstate at which
the evolution (\ref{eq34}) was started is actually untypical; and 3)
this evolution moves eventually the initial state into the manifold
of typical microstates.
It is believed that such features of the stochastic dynamics
(\ref{eq34})-(\ref{eq35}) are owned also by the deterministic
dynamics of a typical many-particle system.
In particular, it is in the weak or effective sense being clarified in
point 3) above that the ergodic hypothesis of statistical mechanics may
actually be relevant for mechanical systems (and are the most) that are
not strictly ergodic~\cite{Goldstein}.

For a given number $n$ of particles in box 1, the equilibrium probability
density $W_n(u)$ of their total energy $u$ can be calculated exactly for
$w(\{{\bf v}\})\propto\delta(v_1^2+\cdots+v_N^2-U)$, by evaluating the
probability that the number $v_1^2+\cdots+v_n^2$ be comprised in an
interval $(a,b)$, with $a>0$.
Once again, this probability is calculated by transforming to
hyperspherical coordinates:
\begin{eqnarray}
&& P(v_1^2+\cdots+v_n^2\in (a,b))
\nonumber \\
&=& \frac{S_{3n}(1)S_{3(N-n)}(1)\int_{\sqrt{a}}^{\sqrt{b}}{\rm d}r\,r^{3n-1}
\int_0^{+\infty}{\rm d}\rho\,\rho^{3(N-n)-1}
\delta(\rho^2+r^2-U)}
{S_{3N}(1)\int_0^{+\infty}{\rm d}r\,r^{3N-1}\delta(r^2-U)}
\nonumber \\
&=& \frac{\Gamma(3N/2)}{\Gamma(3n/2)\,\Gamma(3(N-n)/2)}\,
U^{-\left( \frac{3N}{2}-1\right) }
\int_a^b{\rm d}u\,u^{\frac{3n}{2}-1}(U-u)^{\frac{3(N-n)}{2}-1}\,.
\label{eq48}
\end{eqnarray}
Hence, the final result:
\begin{equation}
W_n(u)=\frac{\Gamma(3N/2)}{\Gamma(3n/2)\,\Gamma(3(N-n)/2)}\,
U^{-\left( \frac{3N}{2}-1\right) }
u^{\frac{3n}{2}-1}(U-u)^{\frac{3(N-n)}{2}-1}\,,
\label{eq49}
\end{equation}
that is, the variable $u/U$ is {\em beta}-distributed with an average of
$n/N$ and a variance of $(n/N)(1-n/N)/(3N/2+1)$ (which is ${\cal O}(N^{-1})$
for $n={\cal O}(N)$).
Of all $n$-velocity microstates, the fraction of those states whose energy
lies between $u$ and $u+\Delta u$ is $W_n(u)\Delta u$ (for $\Delta u\ll u$).
In particular, the Boltzmann entropy associated with Eq.\,(\ref{eq49}) is,
for $n,N-n={\cal O}(N)\gg 1$:
\begin{equation}
\ln W_n(u)\sim -\frac{3n}{2}\ln\frac{n}{u}-\frac{3(N-n)}{2}\ln\frac{N-n}{U-u}
+\frac{3N}{2}\ln\frac{N}{U}\,,
\label{eq50}
\end{equation}
which, when including also the configurational term (\ref{eq08}) or
(\ref{eq18}), gives back the correct expression of the entropy of the
(monoatomic) ideal gas:
\begin{equation}
\frac{S}{k_B}=N\ln\frac{V}{N}+\frac{3N}{2}\ln\frac{U}{N}\,.
\label{eq51}
\end{equation}

\section{Conclusions}

Simple, yet non trivial, theoretical models hold a prominent place
in our own understanding of the physical reality, since they help in
corroborating in our mind the general abstract principles.
This is especially true for the learning of statistical mechanics,
where the appeal of students to their physical intuition, which is
grounded on every-day experience, is not as easy as for classical
mechanics and therefore the convinced acceptance of basic principles
by them would require a proper mediation.

An example is the hypothesis of equal {\it a priori} probability of
all microstates, which is crucial for getting at the microcanonical
and canonical ensembles.
In this paper, I have introduced a stochastic process of the Ehrenfest
type which, among other things, provides a microscopic justification for
the expression of the thermodynamic entropy of an ideal gas, {\it i.e.},
without relying on any ergodic hypothesis.
Rather, the validity of this hypothesis, at least in an effective sense,
arises automatically from the stochastic dynamics itself.
However, in order to make the asymptotics of the present model solvable
a rather strong assumption was made, {\it i.e.}, that positions and
velocities actually behave as uncorrelated random variables.
This is only justified so long as the two sets of variables relax on very
different time scales, which is a fair assumption only for low-density
gases ({\it i.e.}, for particles undergoing only sporadic encounters).
The proposed derivation of the ideal-gas entropy somehow recalls the
heuristic estimate of the multiplicity by \cite{Baierlein,Reif},
though being definitely more rigorous.

In thermodynamics, the Second Law requires the maximizization of the total
entropy $S$ under the given constraints (here, the total number of particles
$N$ and the total energy $U$ of two ideal gases being in grand-canonical
contact with each other) in order to find the equilibrium state of an
overall isolated system.
In the present model, this very same prescription emerges naturally, when
defining the entropy {\it \`a la} Boltzmann, as the condition upon which
the partition of $N$ and $U$ between the gases be, in the long-time regime,
the (overwhelming for $U,N\gg 1$) most probable.
Hopefully, a discussion of this model with the students can serve to deepen
their comprehension of the hypotheses underlying statistical mechanics.

\newpage
%
%

\newpage
%
%
\begin{center}
\large
FIGURE CAPTIONS
\normalsize
\end{center}
\begin{description}
\item[{\bf Fig.\,1 :}]
Numerical simulation of Eq.\,(\ref{eq34}).
Top: Hystogram of velocity values for particle 1 ($\triangle$, $\Box$,
and $\bigcirc$ correspond to the $x,y$, and $z$ component, respectively).
Here, $N=3$ and $U=0.06$.
After rejecting a total of $10^5$ collisions per particle (CPP) (so as to
sweep away any memory of the initial state), as many as $10^7$ CPP are
produced.
The $p$ value is 0.5, held fixed during the simulation.
Data (in form of frequencies of occurrence) are grouped in bins of
width $\delta v=2\sqrt{U/N}/31$.
After equilibration, the hystogram is updated every 10 CPP.
The full curve is the theoretical, finite-$N$ MB distribution per
single velocity component, which is appreciably different from the
infinite-$N$ limit ({\it i.e.}, the Gaussian
$\sqrt{\kappa/\pi}\exp(-\kappa v^2)$, with $\kappa=3N/(2U)$ -- broken curve).
Bottom: Here is plotted the difference between the hystogram and the
finite-$N$ MB distribution.

\item[{\bf Fig.\,2 :}]
Numerical simulation of Eq.\,(\ref{eq34}).
Top: Hystogram of velocity values for particle 1 (same symbols and notation
as in Fig.\,1).
Now, $N=1000$ and $U=20$ ({\it i.e.}, same $U/N$ as in Fig.\,1).
Initially, the $v_a$ vectors are extracted from a uniform one-particle
distribution having zero average and a variance of $U/(3N)$ (hence the
maximum speed $v_{\rm max}=\sqrt{U/N}$).
Then, velocities are rescaled to fit the chosen $U$ value.
After discarding $10^4$ CPP, a huge number of collisions is performed
($10^6$ per particle, with $p=0.5$).
Similarly to $N=3$, data are grouped in bins of width
$\delta v=2v_{\rm max}/31$ and the hystogram is updated every 10 CPP.
The full curve is the theoretical distribution, that is the Gaussian
$\sqrt{\kappa/\pi}\exp(-\kappa v^2)$, with $\kappa=3N/(2U)$.
Bottom: Difference between the hystogram and the above Gaussian.

\item[{\bf Fig.\,3 :}]
Numerical simulation of Eq.\,(\ref{eq34}).
Top: Particle velocities at the end of the simulation run for $N=1000$
and $U=20$ (see Fig.\,2, caption; same symbols and notation as in Fig.\,1).
The distribution of all-particle velocities at a given time strongly
resembles the same Gaussian as in Fig.\,2 (full curve).
Bottom: Difference between the above hystogram and this Gaussian law
(note the change of scale with respect to Figs.\,1 and 2).
\end{description}
\newpage
%
%
\begin{figure}
\begin{center}
\setlength{\unitlength}{1cm}
\begin{picture}(18,15)(0,0)
\put(-1.5,0){\psfig{file=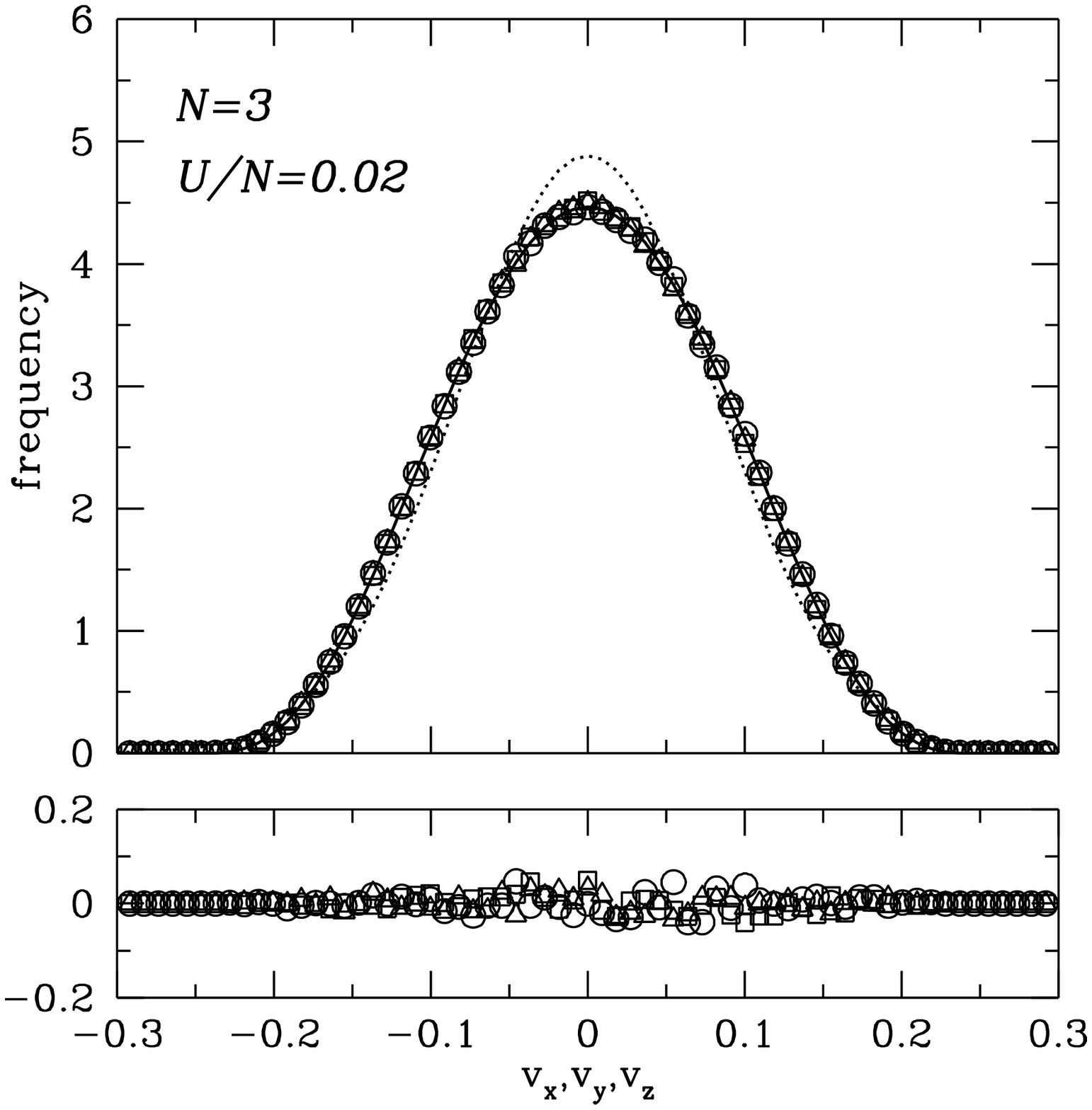,width=18cm,bbllx=0cm}}
\end{picture}
\caption[1]{
}
\end{center}
\end{figure}
%
%
\begin{figure}
\begin{center}
\setlength{\unitlength}{1cm}
\begin{picture}(18,15)(0,0)
\put(-1.5,0){\psfig{file=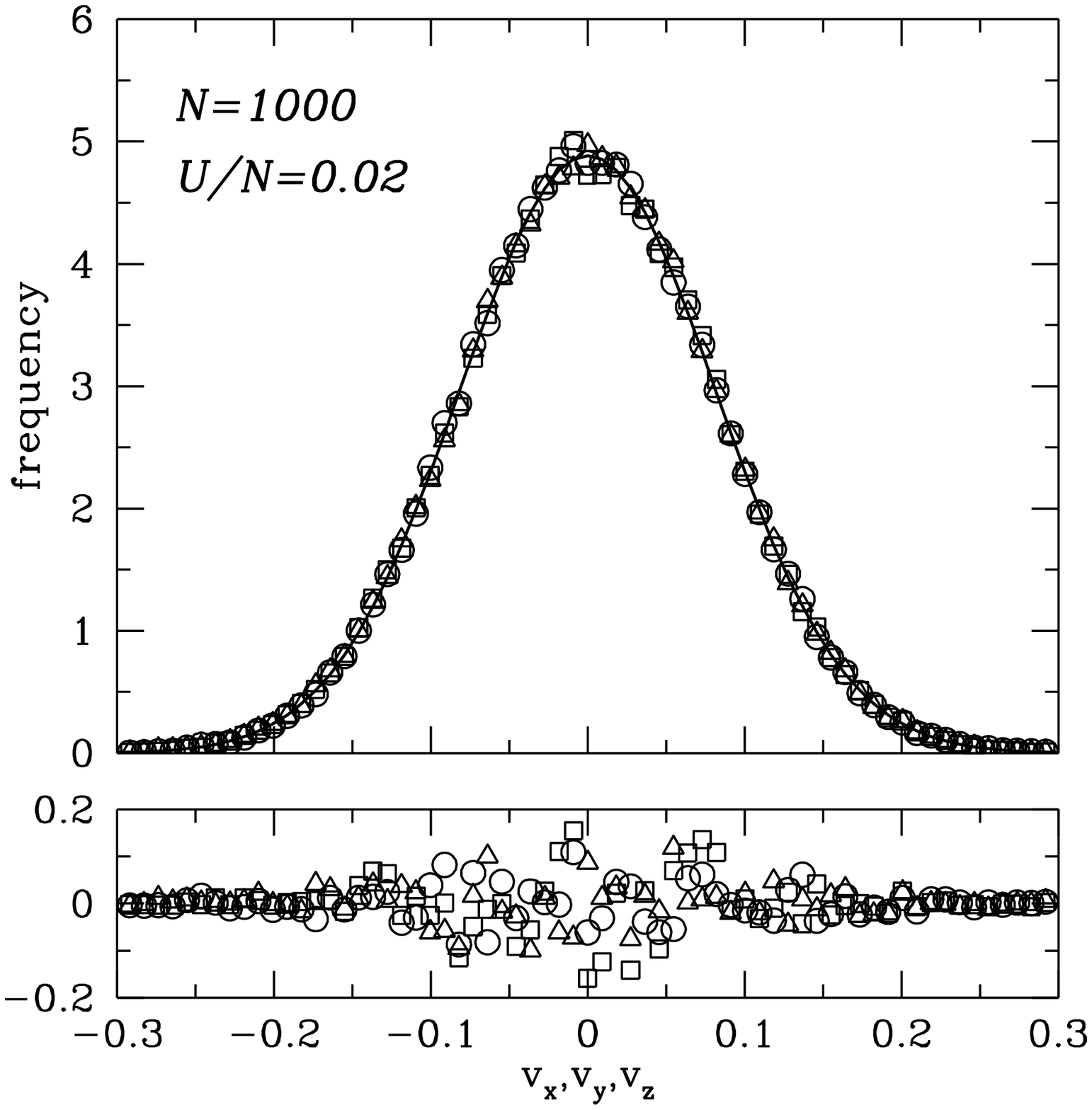,width=18cm,bbllx=0cm}}
\end{picture}
\caption[2]{
}
\end{center}
\end{figure}
%
%
\begin{figure}
\begin{center}
\setlength{\unitlength}{1cm}
\begin{picture}(18,15)(0,0)
\put(-1.5,0){\psfig{file=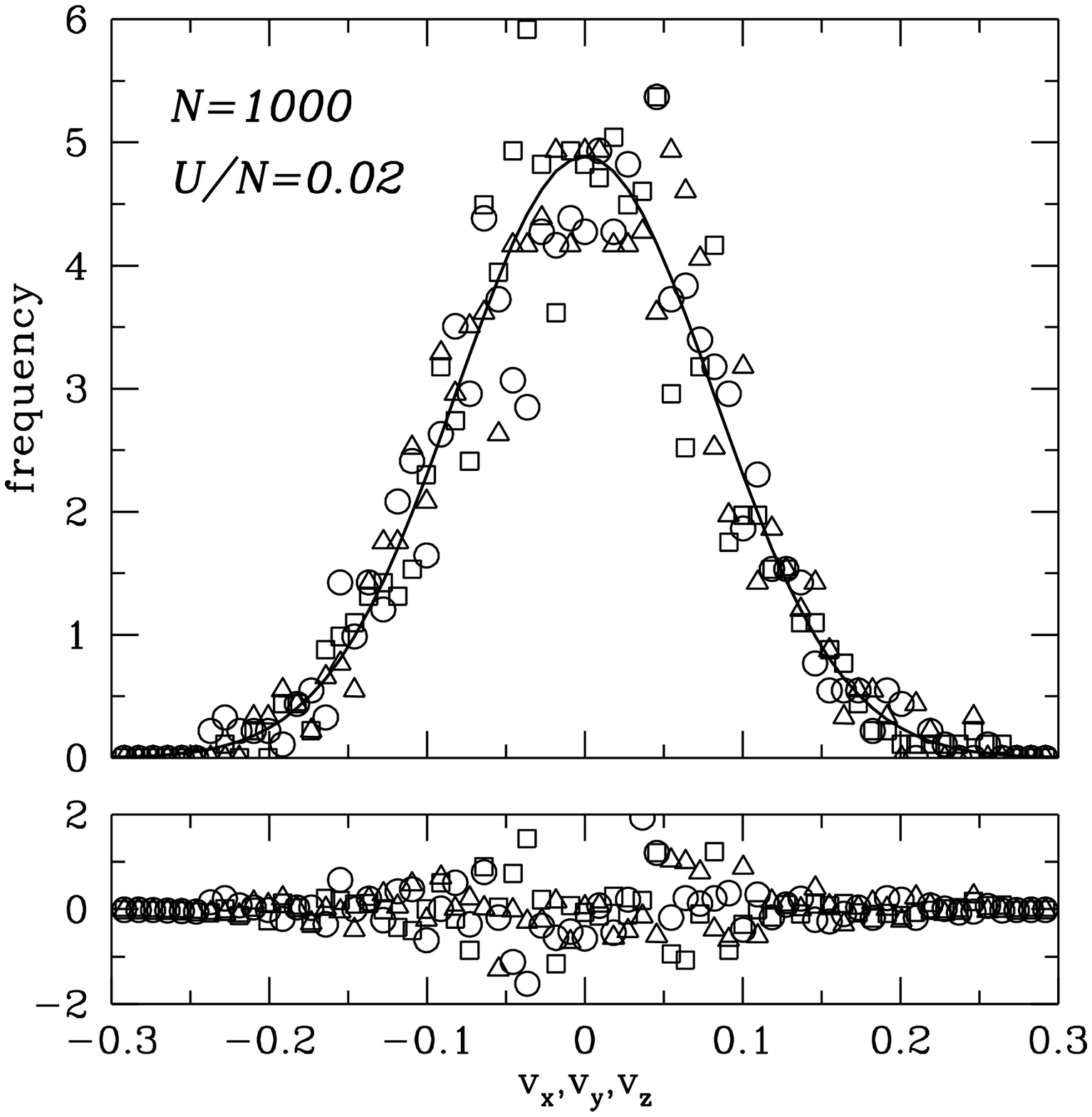,width=18cm,bbllx=0cm}}
\end{picture}
\caption[3]{
}
\end{center}
\end{figure}

\begin{thebibliography}{99}
\bibitem{note}  In order to sample uniformly the 3-dimensional spherical
surface of unit radius, it is necessary to extract $(\theta,\phi)$ pairs
being distributed according to $\frac{1}{2\pi}\times\frac{1}{2}\sin(\theta)$.
In a computer, this is accomplished through the usual generator of
pseudo-random numbers: called $R_{\theta}$ a variate that is uniformly
distributed between 0 and 1, $\Theta={\rm arccos}(1-2R_{\theta})$ is
then distributed according to $(1/2)\sin(\theta)$.
\bibitem{note2}  Hereafter, $U$ is meant to represent the value of the
total kinetic energy in units of $m/2$, $m$ being the particle mass.
\end{thebibliography}
\end{document}